\documentclass[letterpaper, 10pt, conference]{ieeeconf}      

\IEEEoverridecommandlockouts                              

\usepackage{amsmath} 
\usepackage{amssymb}  

\usepackage{longtable}
\usepackage[table,xcdraw]{xcolor}
\usepackage{graphicx}

\usepackage{adjustbox}
\usepackage{multirow}
\usepackage{dblfloatfix}
\usepackage{fancyhdr}
\usepackage{subcaption}

\usepackage{url}

\fancypagestyle{firstpage}{%
    \fancyhf{}
    \fancyfoot[l]{\scriptsize This work was accepted by the International Conference on Human-Machine Systems, Singapore, 2026. \textcopyright 2026 IEEE. Personal use of this material is permitted. Permission from IEEE must be obtained for all other uses, in any current or future media, including reprinting/republishing this material for advertising or promotional purposes, creating new collective works, for resale or redistribution to servers or lists, or reuse of any copyrighted component of this work in other works.}
    
}

\title{\LARGE \bf
Persona-Based Process Design for Assistive Human-Robot Workplaces for Persons with Disabilities
}

\author{Nils Mandischer$^{1}$, Daria Eckert$^{1}$, and Lars Mikelsons$^{1}$
\thanks{*This work was funded by the Bavarian Hightech-Agenda within the AI Production Network Augsburg.}
\thanks{$^{1}$Chair of Mechatronics, University of Augsburg, 86159 Augsburg, Germany. Corresponding: {\tt\small nils.mandischer@uni-a.de}}%
}

\begin{document}
\bstctlcite{IEEEexample:BSTcontrol}

\maketitle
\thispagestyle{empty}
\pagestyle{empty}
\thispagestyle{firstpage}

\begin{abstract}
Human–robot interaction is emerging as an important paradigm for integrating persons with disabilities into the workplace. While these systems can enable individuals to work, their design is mostly personalized, hindering widespread use beyond the individual user. The universal design paradigm is a central pillar of inclusive design, describing usability of systems by all. To incorporate universal design into process design for human-robot workplaces expert knowledge is required that is often not available. To simplify process design of human-robot workplaces, we propose a persona-based design approach. First, typical impairments prevalent in the workforce or particularly relevant for the processes are abstracted into personas with disabilities. The work process is subdivided into sequential actions. For each action and persona, strategies are developed to reach the action goal by a design thinking approach. The resulting actions are ordered by level of robot assistance, i.~e. robot involvement, and implemented in a behavior tree. Therefore, the macro-behavior of the workplace may adapt to individual personas online. We demonstrate the method in a collaborative box folding process with a total of seven personas with disabilities. The persona-based process design shows promising results by generating more comprehensive process strategies while enabling adaptive behavior in the sense of universal design.
\end{abstract}

\section{Introduction}
\label{sec:intro}
Robots become more established as assistive devices for persons with disabilities (PwD) in work settings. However, only few applications have made their way out of the lab into the wild. Often, economic reasons prevent robots from being established in the industry. Nowadays, as assistance for PwD is highly individualized, there are few, if any possibilities to share workplace solutions between employees. Thus, individual assistive devices are expensive, both monetary and in terms of space.

One way to counteract individualization of assistive robots is to make them more adaptable. The universal design paradigm~\cite{Mace.1991} strives for systems that are accessible to all users in an equitable and flexible way. The European Commission's Union of Equality recommends ``\emph{mainstreaming the universal design approach for better accessibility and provision of reasonable accommodation for PwD into all actions.}''~\cite{UnionOfEquality}. Thus, adaptable assistive robots implement this strategy specifically. However, making systems more adaptable requires more diverse users to verify system functionality. To verify such systems with real people is a challenge, as disabilities are complex and participants are not easily acquired. Personas with disabilities can mitigate this challenge~\cite{Schulz.2012}.
In this work, we propose a persona-based approach for the process design for assistive human-robot workplaces interacting with PwD. The approach simplifies the implementation of universal design principles into assistive process design by utilizing design thinking, structuring robot assistance on gradual levels, and, finally, implementing the process directly in a configurable behavior tree.

The paper is structured as follows: Section~\ref{sec:relatedwork} lists related work in process design for people with disabilities and personas thereof. Section~\ref{sec:design} explains our proposed method for process design, incorporating personas with disabilities and design thinking. The section, further, elaborates on the usage of behavior trees as an adaptive task scheduling approach. Lastly, Section~\ref{sec:verification} verifies the method in a collaborative box folding application, before Section~\ref{sec:conclusion} summarizes the work and gives an outlook towards future work on the topic of adaptive robotic devices for inclusion.

\section{Related Work}
\label{sec:relatedwork}

\begin{figure*}[t]
    \centering
    \includegraphics[width=\linewidth]{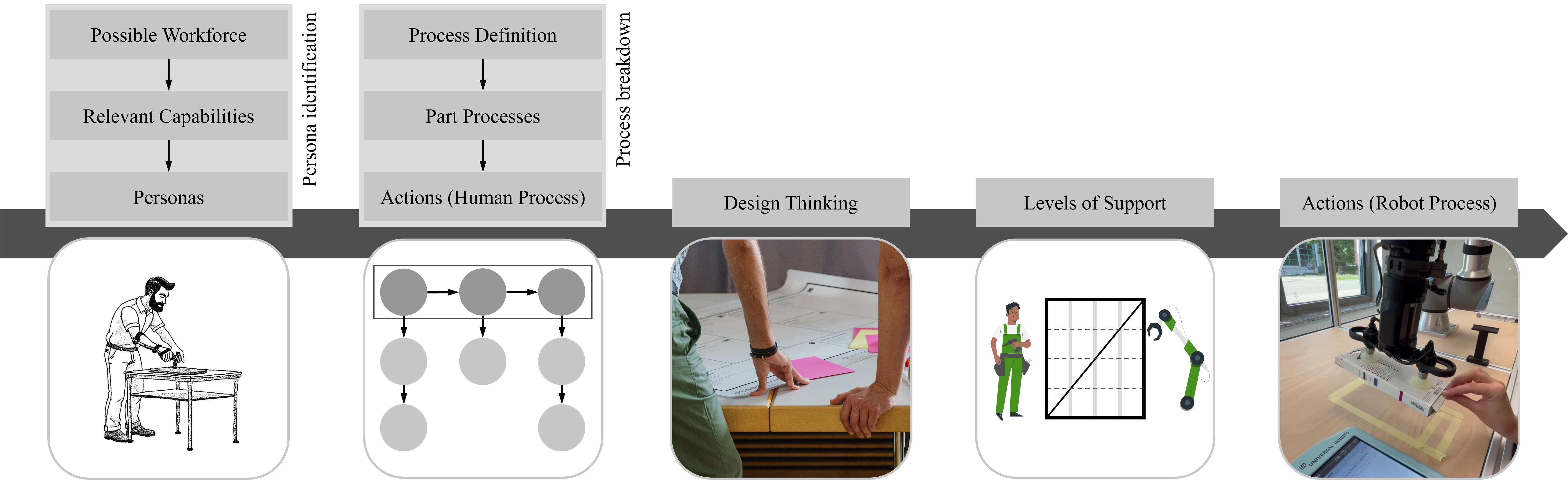}
    \caption{Flowchart of the process design method. The aim of the method is to derive collaborative actions for the robot that complement those for the human. The first two stages may be parallelized. Drawing of PwD is taken from collection~\cite{Mandischer.2025}.}
    \label{fig:flowchart}
\end{figure*}

\subsection{Process Design for Assistive Workplaces}
\label{ssec:sota_process_design}
Multiple authors described studies on robots as assistive devices for the integration of PwD into work~\cite{Weidemann.2022,Drolshagen.2020,Arboleda.2020,Kildal.2019,Graser.2013}, with a recently increasing trajectory. Few works focus on the design of the collaborative process itself, while the majority tackles immaterial or technological challenges. Few applications were industrialized. Ford Motor Company integrated ``Robbie'', a collaborative robot, to support a line worker with reduced mobility~\cite{FordRobbie}. The robot assists the worker's impaired capabilities, while relying on static process design and individualization.
Kildal et al.~\cite{Kildal.2019} derived design requirements for robots assisting persons with cognitive disabilities. However, they arrived at a specialized robot, that highlights electrical connectors via a laser pointer. Kang, Kim, and Chung~\cite{Kang.2008} conducted a survey on the distribution of disabilities to determine a target user group. Based on this, they designed a wheelchair-like assistive device featuring a robot. Understanding the target audience for an assistive workplace plays a pivotal role in their design.

H\"{u}sing et al.~\cite{Husing.2021} proposed RAMB, a software tool to compare capabilities and process requirements for identifying processes for robotic assistance. Later, they embedded RAMB within a product design process based on standard VDI~2221 ``Design of technical products and systems''~\cite{Husing.2024}. Kremer et al.~\cite{Kremer.2018} use a comparison of performative requirements of robot-assisted and non-assisted work processes in sheltered workshops, and quantify the benefit of implementing collaborative robots. Both approaches are based on the German capability assessment and documentation standard IMBA\footnote{\textbf{IMBA}: (translated) Integration of people with disabilities into work.}~\cite{IMBA.2019}. Lottermoser, Y\"{u}cel and, Berger~\cite{Lottermoser.2023} proposed a five-step method for selecting robotic assistance systems. Based on the user’s disabilities, they analyzed existing and required body functions, selected suitable activities, and chose a corresponding interaction system from a database. St\"{o}hr, Schneider, and Henkel~\cite{Stoehr.2018} developed a system model for human-robot collaborative workplaces featuring diverse increments to be considered. Beside the technical system models, environments, sensors, and system integration strategies, they utilized interaction strategies, the type of user disabilities, and robot capabilities to design collaborative work processes. Again, the design decisions are based on the user's and -- in this case robot's -- capabilities, and strategies are chosen accordingly. These works emphasize the importance of understanding users' disabilities and performative capabilities when designing an assistive workplace.

\subsection{Personas in Context of Work and Disability}
\label{ssec:sota_personas}
Personas are fictional but rich representations of groups of humans with similar stereotypic characteristics, originally popularized by Cooper~\cite{Cooper.1999}. Personas guide designers in creating suitable and meaningful user experiences.
Schulz and Skeide Fuglerud~\cite{Schulz.2012} propose a guideline to create personas with disabilities. They emphasize that PwD should participate in the stakeholder group. Further, studying the assistive technologies used by the PwD stakeholders helps in understanding the challenges when interacting with them. This facilitates the collection of factoids on the persona. Finally, the fleshed out personas should feature the disabilities, assistive technologies used, and the consequences on how they interact with their environment. However, Edwards, Sum, and Branham~\cite{Edwards.2020} outline tensions between personas and the complexity of disabilities, as social or self identities of the PwD may interfere with the disability, ultimately changing the subject's behavior.

The usage of personas in design of human-robot interaction (HRI) scenarios is underexplored.
Freitas dos Santos et al.~\cite{DosSantos.2014} and Duque et al.~\cite{Duque.2013} use artificial personas as a means to configure the behavior of robot companions. 
Materna et al.~\cite{Materna.2017} use human personas as a modality to design human-robot collaborative workspaces in virtual reality. There is no work that uses personas in process design for assistive human-robot workplaces.


\section{Persona-Based Process Design}
\label{sec:design}
Our proposed process design approach is based on a four stage sequential process with two parallel stages in the definition of processes and personas. Its aim is to find suitable robot actions that complement human activities in a way that maintains human independence in the work process. Therefore, we want to identify opportunities for robot assistance at gradual levels of intervention intensity. The full process flow is depicted in Figure~\ref{fig:flowchart}.
First, the process is sub-structured. We break down a formal process definition into part processes that require a sequence of actions to complete. Actions act as a gate to further actions in the part process. Failed actions may be repeated until successful. If an action is finally unsuccessful, the part process and, consequently, the process fails. A good process substructure may be generated using Methods-Time Measurement (MTM) portfolio methods, like MTM-HWD (Human Work Design)~\cite{Finsterbusch.2016}. MTM commonly defines actions that are centered around the main process of reaching and bringing items, which is similar to pick \& place processes in robotics. In general, actions should be defined as fundamentally as possible to later avoid challenges when assessing their feasibility in the context of impairments.



In parallel, personas are generated (Section~\ref{ssec:disability_personas}). Both increments, personas and actions, are then assessed in a design thinking approach (Section~\ref{ssec:design_thinking}). From these levels of robot assistance are abstracted (Section~\ref{ssec:levels}). Finally, the levels and robot actions are implemented in a configurable behavior tree (Section~\ref{ssec:behavior_tree}).

\subsection{Personas with Disabilities}
\label{ssec:disability_personas}
When designing adaptive HRI for a diverse user group, it is virtually impossible to have all potential users participate as stakeholders in the persona workshop. We propose to fall back to occupational standards, e.~g. IMBA as used in~\cite{Husing.2021,Kremer.2018}, to accommodate different types of disabilities.
However, basing the personas on occupational standards substitutes the requirement of featuring impaired participants as outlined by Schulz and Skeide Fuglerud~\cite{Schulz.2012} by a requirement to feature at least one participant who is knowledgeable in interpreting these data.

We base the factoids for personas on the occupational assessment standards IMBA and ELA\footnote{\textbf{ELA}: (translated) Assessment of physical performance in work-related activities.}~\cite{Buehne.2020}, where ELA is a variation of Functional Capacity Evaluation (FCE).
An alternative to IMBA adapted for HRI was proposed by H\"{u}sing et al.~\cite{Husing.2021}. In a nutshell, these standards list a variety of physical, sensory, and cognitive capabilities with potential use in work scenarios. As an alternative, the design team may brainstorm a list of capabilities potentially required in manual work. To prevent bias, these should not be specific to the intended work process. Further, the design team should align their understanding of each capability.

From the list of capabilities, skeletons of personas are created who feature impairments in single or multiple capabilities. This helps the designer to move away from concrete disabilities and shift focus to the bodily or cognitive impairments manifested in the work context. Note that there should also be at least one persona that is not subject to any impairments. This persona will later be the reference for the manual process.
Each skeleton is, consequently, elaborated, with a particular focus on the required assistive devices. As a reference, Weidemann, Mandischer, and Corves~\cite{Weidemann.2024} analyzed the accessibility of input and output devices by PwD and provided a selection tool. Lottermoser, Y\"{u}cel, and Berger~\cite{Lottermoser.2023} proposed a similar tool for system-level selection.

\subsection{Design Thinking}
\label{ssec:design_thinking}
After personas and actions have been defined, the next step is to identify which persona can perform which task and in which way a task may be assisted by the robot to reach the action goal as a team. We base this step on design thinking~\cite{Meinel.2011}. Design thinking does not define a rigid framework, but a toolbox of methods to be applied within structured phases. The first steps according to IDEO~\cite{IDEOU} of \emph{framing a question} and \emph{gathering inspiration} from the real world have already been performed by creating the personas and structuring the process. In the following, consecutive stages are listed:

\subsubsection{Synthesize for Action}
The design team reflects on the process and the personas, and converges on a shared understanding of the foundations. The core of this phase is to identify which actions are inaccessible to which personas.

\subsubsection{Generate Ideas}
Next, the design team brainstorms how inaccessible actions may be overcome by robotic assistance. On the one hand, the robot could act collaboratively to assist the PwD, e.~g., by assisting in lifting an item. On the other hand, the full action may be allocated to the robot. This occurs when the persona is entirely unable to engage in collaborative or shared control interaction with the robot. If an action is allocated to the robot, the team must design a robotic sub-process that (a) fulfills the action goal and  (b) allows the human to proceed to the next action.
The aim of this phase is to generate multiple ideas for each combination of action and impairment.

\subsubsection{Make Ideas Tangible}
Modern collaborative robots are easy to program and to be used in prototyping applications. We use the robot directly to create a mock-up for a specific interaction. In case of allocating an action to the robot, often tools are required. In this phase, makeshift and imagined tools are used to explore whether the ideated interaction is feasible. We used 3D printing to prototype tools. An alternative is to use role-playing to enact an interaction with the robot between two humans physically or within descriptive speech~\cite{Valles-Peris.2018}. Role-playing also enables participants to play the role of another person~\cite{Webb.2021} or, in our case, a persona.

\subsubsection{Test to Learn \& Iterate}
Finally, the mock-ups are iterated based on user interaction and feedback. Design thinking ends, when the design team has converged on a strong solution for the specific persona.



\subsection{Levels of Robot Assistance}
\label{ssec:levels}
The levels of robot assistance are a discrete scale that expresses the involvement of the robot in the process. Usually, the number and complexity of robot actions increases with the level of involvement. The scale starts with full manual execution by the human and ends in full automation by the robot, analogous to the SAE levels of autonomous driving or the assistance and automation spectrum by Flemisch et al.~\cite{Flemisch.2012}. In the ``gray area'' between a fully manual and automated process lie gradual shades of collaborative processes, in which actions are either allocated mutually between or shared by the team of human and robot.

The levels of assistance are conditioned on the process and the defined personas. From the persona-individual process plans identified in Section~\ref{ssec:design_thinking}, processes are identified that could be merged. As the variations of the same processes are limited by the number of subsequent actions, the likelihood of finding similar plans is high. In practice, the design process has produced no more than two collaborative part processes. The merged plans are ordered by level of robot involvement, starting with least robot involvement.
\begin{figure}[b]
    \centering
    \includegraphics[width=\linewidth]{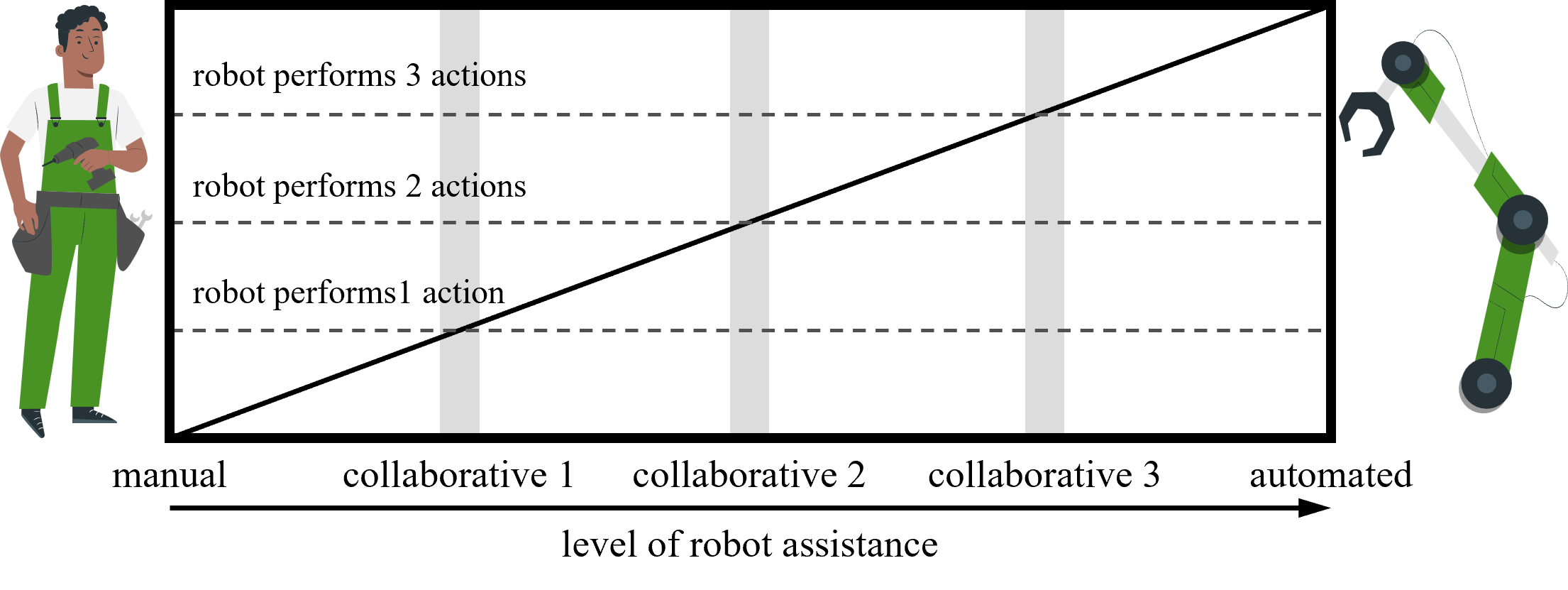}
    \caption{Levels of robot assistance based on the assistance and automation spectrum by Flemisch et al.~\cite{Flemisch.2012}. Here, the robot assists within three tasks with increasing involvement.}
    \label{fig:levels_of_robot_support}
\end{figure}


\begin{figure*}[t]
    \centering
    \includegraphics[width=.95\textwidth]{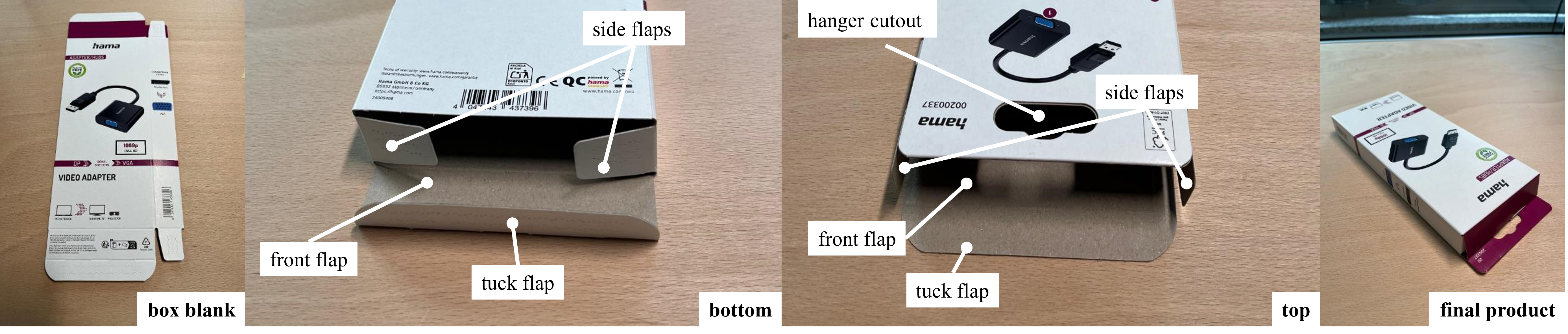}
    \caption{Stages of the box folding process; from left: box blank, opened blank (bottom and top), and folded box.}
    \label{fig:box_stages}
\end{figure*}

\subsection{Behavior Tree}
\label{ssec:behavior_tree}
Human and robot actions may differ significantly. Allocating a human action to the robot may cause the robot to perform multiple sequential actions to fulfill the action goal of the former and let the human proceed within their work process. Actions must, therefore, have a common start and end point for both agents. Collaborative actions are accompanied by the robot waiting for human input (``wait for human''). The final robot actions originate directly from the design thinking process.

To allow the robot to plan with variable personas, the actions are translated into a behavior tree~\cite{Colledanchise.2018}. A behavior tree orders sequential actions within optional sub-trees. Each node requires either all subsequent actions or a single alternative action to conclude. Each persona is associated with an ID. The process is broken down into part processes. For each part process, alternative sequential actions are appended in increasing order of the level of robot assistance. All sequences are parametrized with the personas capable of performing them. This leads to a gradual shift from manual to automated part processes according to Figure~\ref{fig:levels_of_robot_support}. By assigning the persona IDs as conditions, the robot directly chooses the performable sequence. If the real person associated with the persona, e.~g., due to a combination of impairments, is not able to perform the assumed sequence, the behavior tree will wait for a predetermined time and then fall-through to the next higher level of robot assistance. If an action was already accomplished by human or robot, the action goal is flagged as reached, and fall-through processes continue at the correct action. This behavior may continue until the robot performs the part process fully automated. This behavior guarantees that the process continues, even under uncertainty. Note that the robot may not have sufficient capabilities to fully automate certain part processes analogously to the assistance and automation spectrum~\cite{Flemisch.2012}. In this case, the levels of robot assistance may not reach higher levels of collaboration or full automation, and the part process may still fail.

By the described configurable behavior tree with fall-through behavior, a macro-behavior may be chosen for a certain user group, allowing robot workplaces to adapt to the needs of the PwD interacting with them. By using a behavior tree as a process scheduler, the individual robot actions may be statically programmed on the robot controller, enabling the use of inherent safety functions and real-time guarantees.

\section{Verification}
\label{sec:verification}
To verify the approach, we tackle a box folding process provided by Hama GmbH \& Co. KG (see Figure~\ref{fig:box_stages}).
Currently, the task is subject to primary labor market and sheltered workshops, featuring people with diverse impairments. The process consists of eight sequential part processes (compare Figure~\ref{sfig:part_processes} at the end of the paper):
\begin{enumerate}
    \item Unfold box blank
    \item Fold main and side flaps (bottom)
    \item Apply glue (bottom)
    \item Plug main flap over side flaps (bottom)
    \item Fill box
    \item[6-8)] Repeat part processes 2-4 (top)
\end{enumerate}
To not require additional complex and unsafe tooling, we ignored steps 3 and 7, focusing on folding and packing.

\begin{figure}[t]
    \centering
    \includegraphics[width=.95\linewidth]{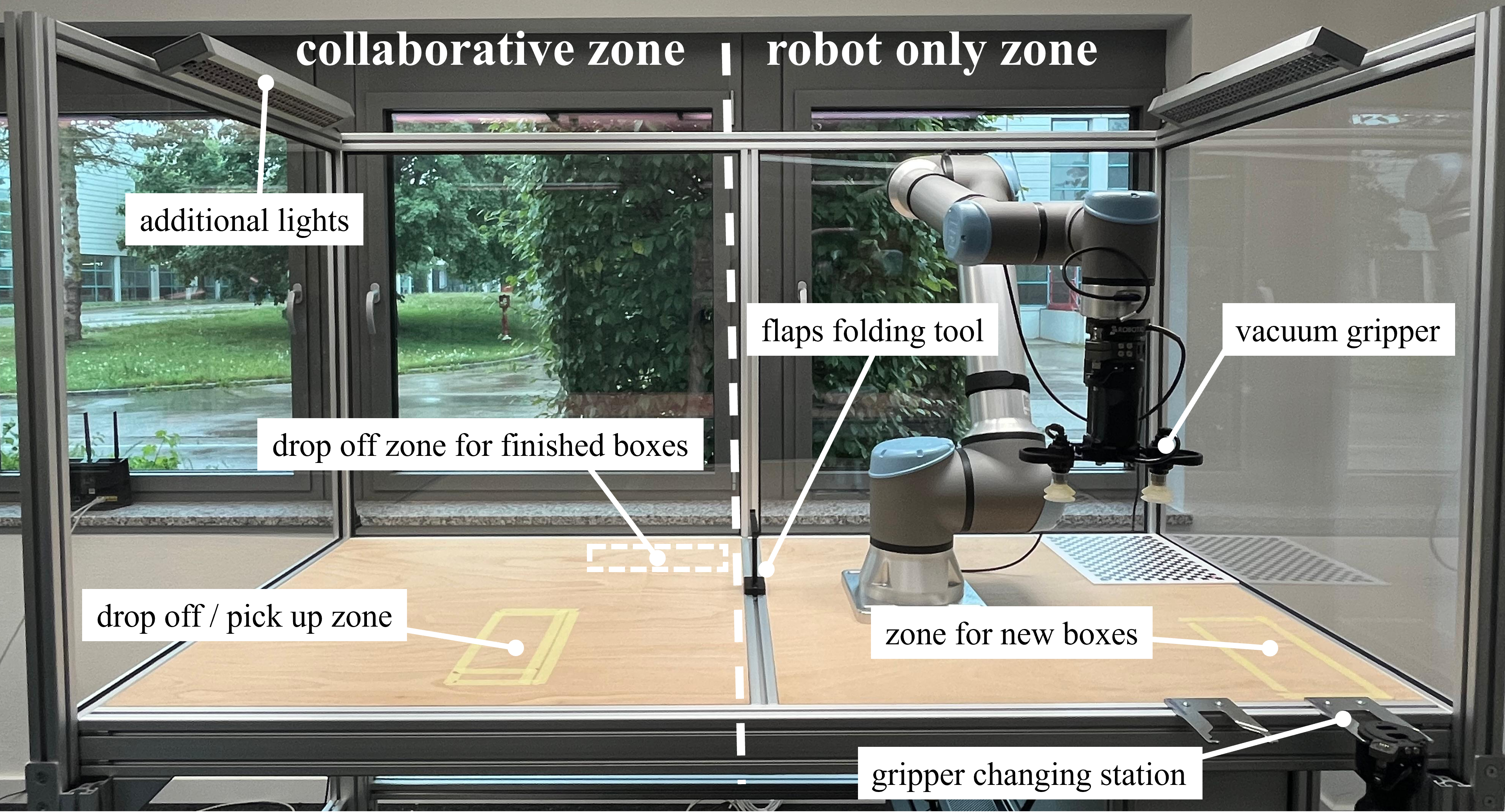}
    \caption{Workplace layout including folding tool ideated during design thinking. Touchscreen not displayed.}
    \label{fig:workplace}
\end{figure}

The layout and hardware selection of the human-robot workstation was finalized beforehand, although additional tools may still be added as part of the design thinking process. The robot (Universal Robots UR10e) is equipped with a wrist camera (Robotiq Wrist Camera) and switchable vacuum (Robotiq EPick) and finger grippers (Robotiq Hand-E). The workspace is divided into a collaborative and a robot only zone (see Figure~\ref{fig:workplace}). Tools usable by both may be placed at the border. The robot communicates instructions via a touch screen. The human is required to acknowledge the completion of actions allocated to them by pressing a virtual button. This interface was chosen to reduce uncertainties of sensor- or reasoning-based state observation. Hardware buttons are available but were not used for the verification.

\subsection{Personas}
\begin{figure}[t]
    \centering
    \includegraphics[width=.95\linewidth]{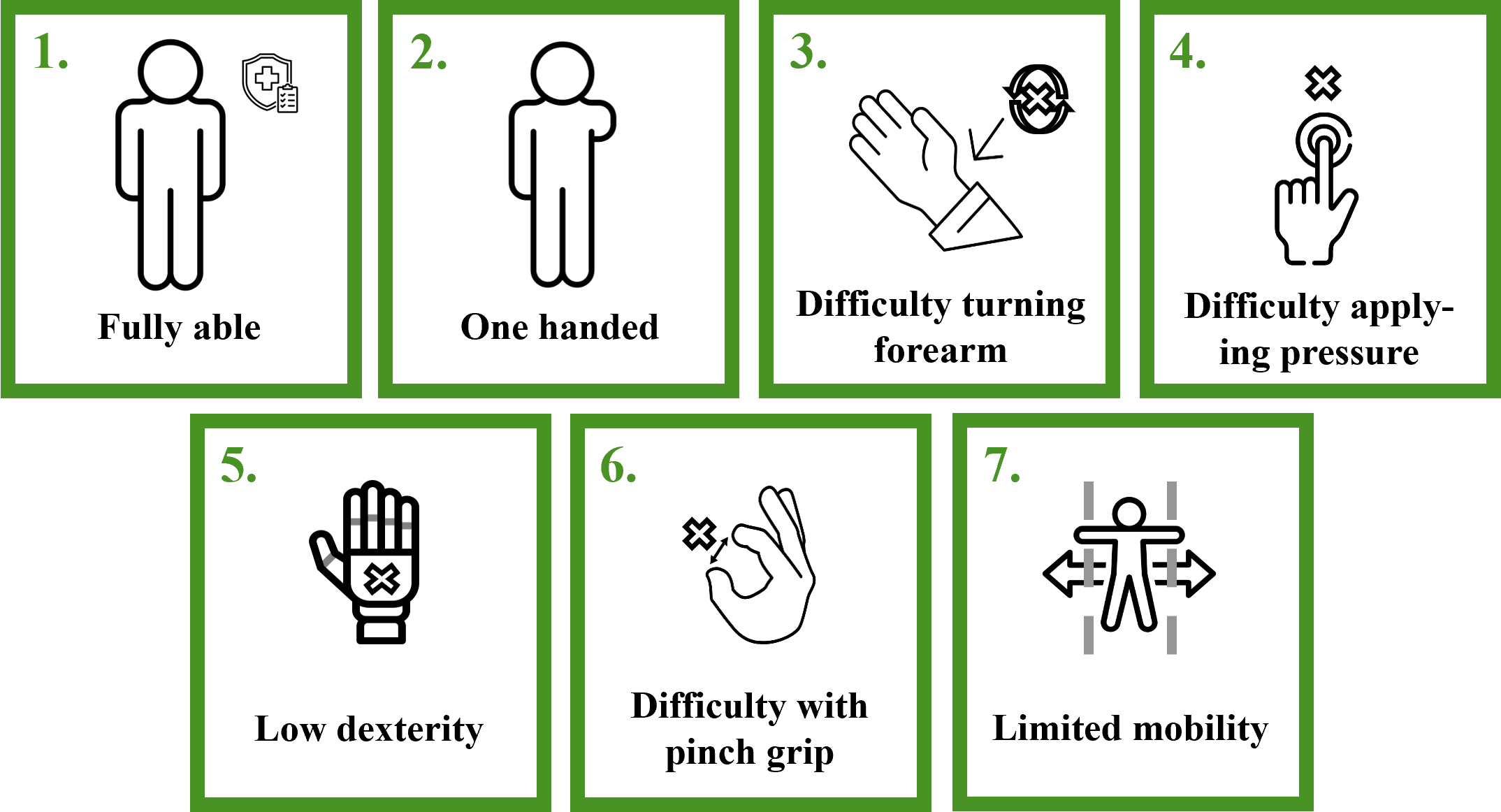}
    \caption{Impairments of the seven personas identified for the box folding process.}
    \label{fig:personas}
\end{figure}

First, we analyzed which capabilities are required to perform the box folding process. These are mainly the capabilities associated with arm, hand, and torso movements. We based the discussion of capabilities on the IMBA standard and identified six categories of potential impairments:
\begin{itemize}
    \item Reaching
    \item Turning of lower arm and/or hand
    \item Fist and/or pinch grip
    \item Applying pressure with finger and/or hand
    \item Finger and/or hand dexterity
    \item Neck and/or trunk rotation
\end{itemize}
From the resulting impairments depicted in Figure~\ref{fig:personas} we generated seven personas. The personas depict a wide variety of users with isolated and combined impairments.

\subsection{Design Thinking}
Our design team featured the minimum size of two people, one of whom was certified in the IMBA standard. First, open questions regarding the definition of capabilities in IMBA were resolved. Second, we brainstormed which part processes were inaccessible for certain personas. This often involved initially assuming more part processes were inaccessible and then discussing whether the impairment was truly so limiting. Often this resulted in less actions to be allocated to the robot. In few cases, the inaccessibility was removed by simply re-arranging item or tool positions.

In \emph{Making Ideas Tangible}, we used a mixture of role-playing and mock-ups. First, we role-played potential robot movements using the robot and hand/arm movements, and tools by using fingers or office supplies like rulers. For instance, initially we considered unfolding the box bank by inserting and pressing the blank against a thin long object (here: a ruler). Second, we built tool mock-ups using 3D printing. This resulted in three tools. The mock-ups showed that the ``ruler'' approach was infeasible. Finally, we re-iterated the tools until we converged on feasible solutions for steps 2, 4, 6, and 8. We were unable to find an efficient automated part process for step~1.

Instead of using full automation in step~1, we reverted to \emph{Generate Ideas} for step~1 alone. We observed that the step was not accessible for personas 2, 5, 6, and 7. By observing how a member of the design team unfolded the blank, we identified that it is sufficient to just lift and hold the blank at an appropriate pose to allow unfolding with one hand and limited gripping capabilities. In fact, while reiterating concepts, many solutions shifted from complex to very simple. One important observation is that very simple robot actions, like holding the box at an ergonomic pose, often resulted in sufficient collaborative actions that allowed the process to proceed.

\subsection{Behavior Tree}
\begin{figure*}[t]
    \centering
    \begin{subfigure}[t]{.95\textwidth}
        \centering
        \includegraphics[width=\textwidth]{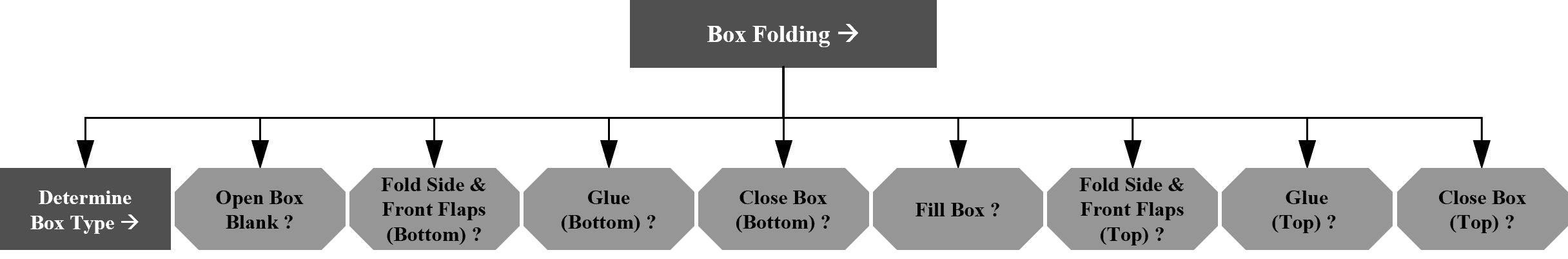}
        \caption{Process structured into sequential part processes. Question marks indicate that conditions are required to proceed to the next horizontal node in the behavior tree.}
        \label{sfig:part_processes}
    \end{subfigure}
    \begin{subfigure}[t]{.49\textwidth}
        \centering
        \raisebox{17mm}{\includegraphics[width=.85\textwidth]{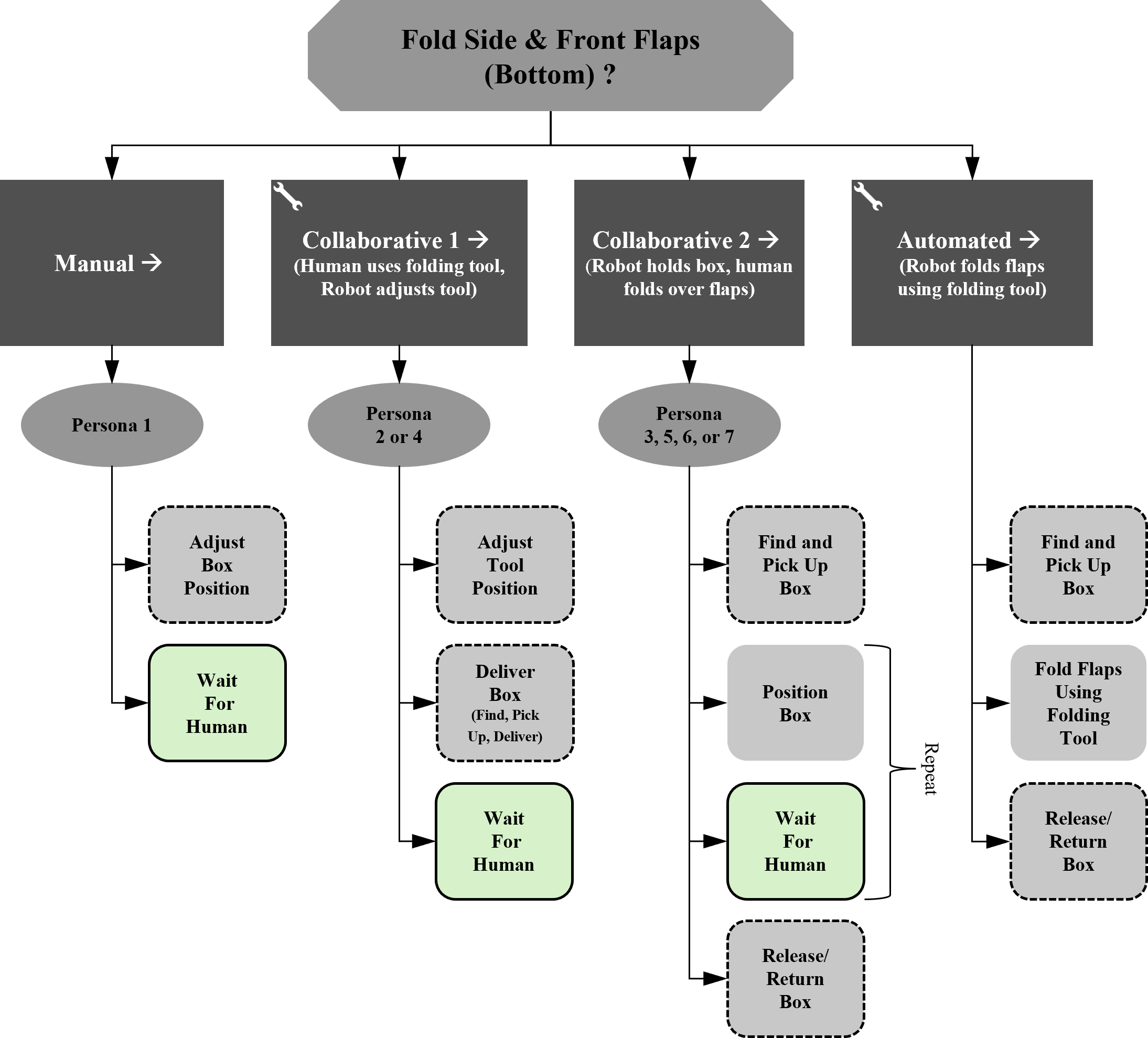}}
        \caption{Flap folding part process structured into four optional action sequences, including two collaborative strategies.}
        \label{sfig:folding_flaps}
    \end{subfigure}
    \begin{subfigure}[t]{.49\textwidth}
        \centering
        \includegraphics[width=\textwidth]{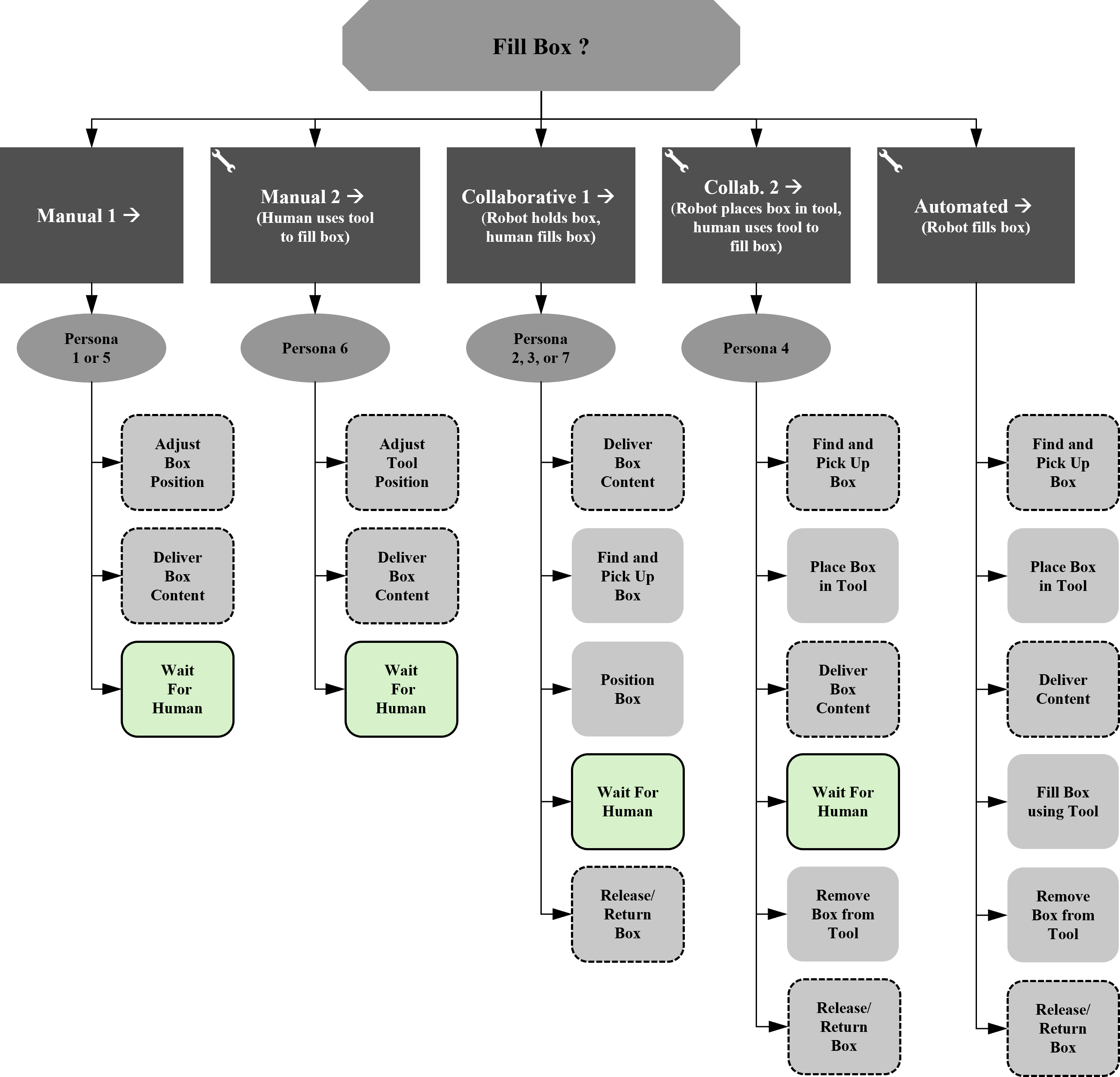}
        \caption{Box filling part process structured into five optional action sequences, including two manual and collaborative strategies. The second manual strategy allows the usage of a tool for box filling without robot assistance.}
        \label{sfig:filling_box}
    \end{subfigure}
    \caption{Behavior tree generated for the box folding process. Two example part processes are depicted in detail. If a strategy is unsuccessful, the tree falls through to the next strategy to the right until the robot takes over the full part process in ``automated''. Dotted lines indicate optional actions depending on the predecessor action, e.~g., if the robot has already gripped the box in the prior action, it does not have to find and grab it.}
    \label{fig:behavior_tree}
\end{figure*}

From the converged plans, we constructed a behavior tree using py\_trees~\cite{PyTreesGit} and ROS~2. The top level of the behavior tree and two detailed partial processes are depicted in Figure~\ref{fig:behavior_tree} (end of paper). In all part processes we also considered the delivery of process items and the handover between part processes. We, further, defined an intermediate robot configuration to simplify transitions between part processes.

The individual actions were programmed directly on an Universal Robots UR10e using visual programming. The actions are then triggered through the behavior tree via ROS~2 interfaces. By using this structure, it is possible for non-experts to modify the actions within the much simpler block language than in, e.~g., Python. In application, a robot expert could design the comprehensive and exhaustive behavior tree, while the non-expert user could create, modify, and maintain the individual robot actions.

\subsection{Learnings}
\begin{figure*}[t]
    \centering
    \includegraphics[width=.92\textwidth]{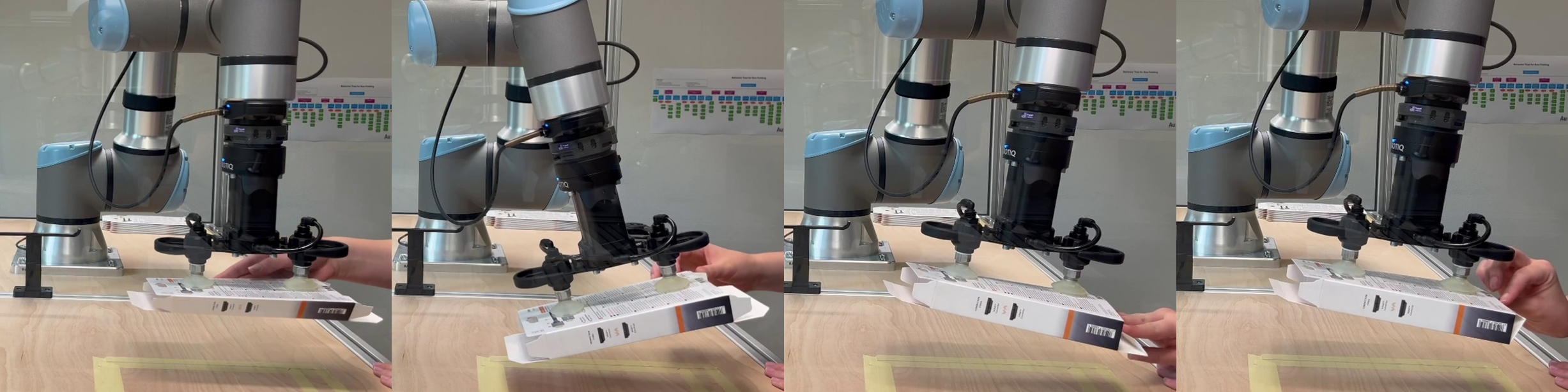}
    \caption{Snippets from assisting persona 7; from left: Collaborative box opening, collaborative stategy~2 in flap folding, and collaborative closing. The robot positions the box close to the worker, and adjusts ergonomic poses.}
    \label{fig:robot_trial}
\end{figure*}

We observed that the structured process simplified the iterations towards the final process design and helped to uncover leaner assistive strategies. The resulting behavior tree differs significantly from the initial structure that contained a wider variety of different collaborative strategies, required more tools, and featured, in general, more complex actions. Within the design thinking phase, \emph{Making Ideas Tangible} was particularly useful to design leaner collaborative strategies and to uncover potential for re-use of tools. For instance, Figure~\ref{fig:robot_trial} depicts that the robot can compensate for an impaired arm by holding the box in a static ergonomic pose. While seemingly trivial, this was only uncovered during experimentation with the live robot. Initial strategies had featured a jaw gripper and more dynamic movements of the robot that were difficult to coordinate with the human.

\section{Conclusion} 
\label{sec:conclusion}
In summary, we presented a persona-based process design method for assistive human-robot workplaces. The method implements the universal design paradigm by enabling alternative process strategies depending on the users' impairments. The method uses personas with disabilities in a design thinking approach, and embeds the results within a configurable behavior tree. The levels of robot involvement are ordered in levels of robot assistance that help in designing fall-through behaviors and autonomy of the robot system. We verified the method in the application of collaborative box folding within a small design team. The results show that the method produces comprehensive and user-centered process strategies.

By applying our method, assistive human-robot processes and workplaces may be created that cater to the universal design principles, allowing a multitude of diverse users while lowering cost and space requirements, and integrating in existing environments and hardware. By this, we contribute to lowering the barrier to implement assistive human-robot workplaces and aim to bridge the gap between lab experiments and widespread industrial applications.

In future work, generative AI should be explored to guide the generation of personas with disabilities. Further, first studies described the usage of ChatGPT in design thinking itself~\cite{Chen.2024,Asadi.2023}. Using LLMs within our approach may make it more accessible to design teams less experienced with disabilities and universal design.




\section*{ACKNOWLEDGMENT}
We like to thank Hama GmbH \& Co. KG for providing the box blanks, Luka Eberle for implementing the behavior tree, and the University of Augsburg for financial support of the Young Researchers Travel Scholarship Program.

\bibliographystyle{IEEEtran}
\bibliography{references}

\end{document}